\documentclass[12pt]{article}
\usepackage{epsfig}
\usepackage{amsfonts}
\begin{document}
\begin{titlepage}
\begin{center}

{\Large Stretched exponentials from superstatistics}

\vspace{2.cm} {\bf Christian Beck}

\hspace{2cm}

School of Mathematical Sciences, Queen Mary, University of
London, Mile End Road, London E1 4NS, UK

\vspace{2cm}

\end{center}

\abstract{ Distributions exhibiting fat tails occur frequently in
many different areas of science. A dynamical reason for fat tails
can be a so-called superstatistics, where one has a superposition
of local Gaussians whose variance fluctuates on a rather large
spatio-temporal scale. After briefly reviewing this concept, we
explore in more detail a class of superstatistics that hasn't
been subject of many investigations so far, namely
superstatistics for which a suitable power $\beta^\eta$ of the
local inverse temperature $\beta$ is $\chi^2$-distributed. We
show that $\eta >0$ leads to power law distributions, while $\eta
<0$ leads to stretched exponentials. The special case $\eta =1$
corresponds to Tsallis statistics and the special case $\eta =-1$
to exponential statistics of the square root of energy. Possible
applications for granular media and hydrodynamic turbulence are
discussed.}

\end{titlepage}

\section{Introduction}

Nonextensive statistical mechanics \cite{tsa1,tsa2,tsa3,abe} was
originally developed as an equilibrium formalism, but most
physical applications of this formalism actually occur for typical
nonequilibrium situations. Sometimes these nonequilibrium
situations are described by a fluctuating parameter $\beta$,
which may, for example, be the inverse temperature. Alternatively,
$\beta$ may be an effective friction constant, a changing mass
parameter, a changing amplitude of Gaussian white noise, a
fluctuating local energy dissipation or simply a local inverse
variance parameter extracted from a time series. The fluctuations
of $\beta$ induce a superposition of different statistics on
different time scales, in short a superstatistics
\cite{beck-cohen, boltzmann-m, beck-su, touchette-beck, yamano,
plastino, ryazanov, prl, wilk, souza, souza2, luczka, garcia,
sattin, sattin2, erice, chavanis, haenggi, grigolini}. The
stationary probability distributions of superstatistical systems
typically exhibit much broader tails than a Gaussian
distribution. These tails can decay e.g. with a power law, or as
a stretched exponential, or in an even more complicated way
\cite{touchette-beck}. Which type of tails are produced depends
on the probability distribution $f(\beta)$ of the parameter
$\beta$. Recent applications of the superstatistics concept
include a variety of physical systems. Examples are Lagrangian
\cite{reynolds, beck03, boden, aringazin} and Eulerian turbulence
\cite{beck-physica-d, jung-swinney, BCS}, defect turbulence
\cite{daniels}, atmospheric turbulence \cite{rapisarda, rap2},
cosmic ray statistics \cite{cosmic}, solar flares \cite{maya},
solar wind statistics \cite{burlaga}, networks \cite{abe-turner,
hasegawa}, random matrix theory \cite{abul-magd}, and
mathematical finance \cite{bouchard, ausloos, hasegawa2}.

If $\beta$ is distributed according to a particular probability
distribution, the $\chi^2$-distribution, then the corresponding
marginal stationary distributions of the superstatistical system
obtained by integrating over all $\beta$ are given by the
generalized canonical distributions of nonextensive statistical
mechanics \cite{tsa1,tsa2,tsa3,abe}. For other distributions of
the intensive parameter $\beta$, one ends up with more
complicated statistics.

In this paper, after briefly reviewing the superstatistics
concept, we explore a rather general case which may be of
relevance to many practical applications. We consider the case of
a superstatistics where $\beta^\eta$, i.e. $\beta$ to some power
$\eta$, is $\chi^2$-distributed, where $\eta$ is some arbitrary
parameter. The case $\eta=1$ is fully understood: It leads to
Tsallis statistics and asymptotic power-law decay of the marginal
distributions obtained by integrating over all $\beta$. However,
the other values of $\eta$ are interesting as well, and will be
explored in more detail here. For general $\eta
>0$ we obtain asymptotic power law decay, though the resulting
statistics is slightly different from Tsallis statistics (only
$\eta =1$ leads exactly to Tsallis statistics). For $\eta <0$ one
obtains tails that asymptotically decay as stretched
exponentials. The special case $\eta =-1$ corresponds to
exponential tails of the square root of energy. We will provide
some arguments (based on the ordinary Central Limit Theorem) why
nonequilibrium systems with many degrees of freedom often lead to
one of the superstatistics described above.

\section{Various types of superstatistics}

It is well known that for the canonical ensemble the probability
to observe a state with energy $E$ is given by
\begin{equation}
p(E)=\frac{1}{Z(\beta)}\rho(E) e^{-\beta E}.
\end{equation}
$e^{-\beta E}$ is the Boltzmann factor, $\rho(E)$ is the density
of states and $Z(\beta)$ is the normalization constant of $\rho
(E)e^{-\beta E}$. For superstatistical systems, one generalizes
this approach by assuming that $\beta$ is a random variable as
well. Indeed, a driven nonequilibrium system is often
inhomogeneous and consist of many spatial cells with different
values of $\beta$ in each cell. The cell size is effectively
given by the correlation length of the continuously varying
$\beta$-field. If we assume that each cell reaches local
equilibrium very fast, i.e.\ the associated relaxation time is
short, then in the long-term run the stationary probability
distributions $p(E)$ arise as the following mixture of Boltzmann
factors:
\begin{equation}
p(E)=\int_0^\infty f(\beta)  \frac{1}{Z(\beta)} \rho(E) e^{-\beta
E}d\beta  \label{ppp}
\end{equation}
Without restriction of generality, we may absorbe the factor
$1/Z(\beta)$ into the function $f(\beta)$, i.e. define $\tilde{f}
(\beta)=f(\beta )/Z(\beta)$ and rename $\tilde{f}\to f$. Also,
for reasons of simplicity we may just assume $\rho(E)=1$, keeping
in mind that the most general case may correspond to a different
density of states. The result is an effective distribution
\begin{equation}
p(E)\sim \int_0^\infty f(\beta)e^{-\beta E}d\beta \label{pe}
\end{equation}
given essentially by the Laplace transform of $f(\beta)$.

The simplest dynamical example of a superstatistical system is a
Brownian particle of mass $m$ moving through a changing
environment in $d$ dimensions. For its velocity $\vec{v}$ one has
the local Langevin equation
\begin{equation}
\dot{\vec{v}}=-\gamma \vec{v} + \sigma \vec{L}(t) \label{lange}
\end{equation}
($\vec{L}(t)$: $d$-dimensional Gaussian white noise, $\gamma$:
friction constant, $\sigma$: strength of noise) which becomes
superstatistical if the parameter $\beta :=\frac{2}{m}
\frac{\gamma}{\sigma^2}$ is regarded as a random variable as well.
In a fluctuating environment $\beta$ may vary from cell to cell on
a large spatio-temporal scale. Of course, for this example the
energy $E$ is just kinetic enery $E=\frac{1}{2}m\vec{v}^2$. In
each cell of constant $\beta$ the local stationary velocity
distribution is Gaussian,
\begin{equation}
p(\vec{v}|\beta)=\left( \frac{\beta}{2\pi}\right)^{d/2}
e^{-\frac{1}{2}\beta m\vec{v}^2},
\end{equation}
provided the relaxation time $\gamma^{-1}$ is small enough as
compared to the changes of $\beta$. The marginal distribution
describing the long-time behaviour of the particle
\begin{equation}
p(\vec{v})=\int_0^\infty f(\beta)p(\vec{v}|\beta)d\beta
\label{margi}
\end{equation}
exhibits fat tails. The large-$|v|$ tails of the distribution
(\ref{margi}) depend on the behaviour of $f(\beta )$ for $\beta
\to 0$ \cite{touchette-beck}. Different superstatistical models
corresponding to different $f(\beta)$: The function $f$ is
determined by the environmental dynamics of the nonequilibrium
system under consideration.

Consider, for example,
a simple case where there are $n$ independent Gaussian random variables $X_1,
\ldots , X_n$ underlying the dynamics of $\beta$ in an additive
way.
$\beta$ needs to be positive and
a positive $\beta$ is obtained by squaring
these Gaussian random variables. The probability distribution
of a random variable that is the sum of
squared Gaussian random variables
$\beta=\sum_{i=1}^nX_i^2$ is well known in statistics:
It is the $\chi^2$-distribution of degree $n$,
i.e. the probability density $f(\beta)$ is given by
\begin{equation}
f(\beta )=\frac 1{\Gamma (\frac n2)}\left( \frac n{2\beta _0}\right)
^{n/2}\beta ^{n/2-1}e^{-\frac{n\beta }{2\beta _0}},  \label{chi2}
\end{equation}
where $\beta_0$ is the average of $\beta$. In this case the
integration over the fluctuating $\beta$ can be explicitly done
and one obtains \cite{prl}
\begin{equation}
p(\vec{v}) \sim \frac{1}{(1+\tilde{\beta}
(q-1)\frac{1}{2}m\vec{v}^2)^{\frac{1}{q-1}}}
\end{equation}
with
\begin{equation}
q=1+\frac{2}{n+d}
\end{equation}
and
\begin{equation}
\tilde{\beta}=\frac{2\beta_0}{2-(q-1)d}.
\end{equation}
These types of distributions, so-called $q$-exponentials, are
relevant for nonextensive statistical mechanics
\cite{tsa1,tsa2,tsa3,abe}.

For other systems the random variable $\beta$ may be generated by
multiplicative random processes. In such cases one typically ends
up with a log-normally distributed $\beta$, i.e.\  the
probability density is given by
\begin{equation}
f(\beta )=\frac{1}{\sqrt{2\pi}s\beta}
\exp \left\{ \frac{-(\ln \frac{\beta}{\mu})^2}{2s^2}\right\},
\label{logno}
\end{equation}
where $\mu$ and $s$ are parameters. The corresponding
superstatistics is relevant for turbulent systems
\cite{beck-physica-d, jung-swinney, BCS}. The integration over
$\beta$ cannot be done analytically in this case.

\section{Generalization}

Let us now generalize the additive case. For more general
nonequilibrium situations, it may not $\beta$ itself that is
$\chi^2$-distributed, but some suitable power of $\beta$, say
$y:=\beta^\eta$, where $\eta$ is a parameter. What the relevant
power $\eta$ is may depend on the physical nature of the
nonequilibrium system under consideration and the physical meaning
of $\beta$. $\eta$ can depend on the spatial dimensions of the
experiment under consideration, its boundary conditions, the
dissipation mechanism, the nature of the driving forces, etc.

To construct a microscopic model for $y=\beta^\eta$, we may assume
that there are many (nearly) independent microscopic random
variables $\xi_j$, $j=1,\ldots , J$, contributing to $y$ in an
additive way. For large $J$ their rescaled sum
$\frac{1}{\sqrt{J}}\sum_{j=1}^J\xi_j$ will approach a Gaussian
random variable $X_1$ due to the Central Limit Theorem. In total,
there can be many different random variables consisting of
microscopic random variables, i.e., we have $n$ Gaussian random
variables $X_1,\ldots ,X_n$ due to various relevant degrees of
freedom in the system. $y$ must be positive. A positive value is
obtained by squaring the $X_i$. By construction, the resulting
$y=\sum_{i=1}^nX_i^2$ is $\chi^2$-distributed with degree $n$,
i.e.\ one has the probability density
\begin{equation}
f_y(y)=\frac 1{\Gamma (\frac{n}{2})}\left( \frac n{2y_0}\right)
^{n/2}y^{n/2-1}e^{-\frac{ny }{2y_0}},  \label{chi22}
\end{equation}
where $y_0$ is the average of $y$.

A priori, all kinds of values of $\eta$ are possible. For example,
$\eta=-1$ means that the temperature $\beta^{-1}$ is the relevant
quantity that can be represented as a sum of several squared
Gaussian random variables. Another case may be that the local
standard deviation of the Gaussians is $\chi^2$-distributed,
corresponding to the case $\eta=-\frac{1}{2}$, and so on. Each
choice of $\eta$ implies a different probability distribution of
$\beta$. The resulting probability density $f_\beta (\beta)$ can
simply be obtained by employing a transformation from the
$\chi^2$-distributed random variable $y=\beta^\eta$ to the
original random variable $\beta$ via $f_y(y)dy=f_\beta
(\beta)d\beta$. One obtains
\begin{equation}
f_\beta (\beta)=\frac{|\eta|}{\Gamma (\frac{n}{2})} \left(
\frac{n}{2y_0} \right)^{\frac{n}{2}} \beta ^{\eta
\frac{n}{2}-1}e^{-\frac{n}{2y_0}\beta^\eta} \label{chichi},
\end{equation}
where $y_0=\langle y\rangle=\langle \beta^\eta \rangle$ is the
average of $y$. As we shall see in the next section, for $\eta
>0$ this generates asymptotic power laws for $p(E)$, whereas the case $\eta
<0$ generates stretched exponentials. Both cases can be
physically relevant.

\section{Asymptotic behaviour}

Let us study the marginal probability distributions $p(E)$ as
given by eq.~(\ref{pe}) for our class of inverse temperature
distributions (\ref{chichi}). In particular, we are interested in
the behaviour of $p(E)$ for large $E$, i.e.\ the shape of the
tails.

Two important cases have to be distinguished, $\eta >0$ and $\eta
<0$. As has been shown in \cite{touchette-beck}, the large-energy
tails of $p(E)$ are determined by the behaviour of the function
$f(\beta)$ for $\beta \to 0$. For $\eta
>0$ eq.~(\ref{chichi}) yields the small-$\beta$ asymptotics
\begin{equation}
f_\beta (\beta) \sim \beta^{\eta \frac{n}{2}-1},
\end{equation}
since the exponential term $\exp(-\frac{n}{2y_0}\beta^\eta)$ in
eq.~(\ref{chichi}) just approaches 1. According to the general
results of \cite{touchette-beck} this implies
\begin{equation}
p(E) \sim E^{-\eta \frac{n}{2}}
\end{equation}
for $E\to \infty$. This means one obtains an asymptotic power-law
decay in $E$ with the exponent $\eta \frac{n}{2}$. For example,
if the energy is just kinetic energy, $E= \frac{1}{2}mv^2$, then
$p(v)$ decays with tails proportional to $|v|^{-\eta n}$. This
reminds us of Tsallis statistics and nonextensive statistical
mechanics \cite{tsa1,tsa2,tsa3,abe}, with an entropic index $q$
given by
\begin{equation}
\frac{2}{q-1} = \eta n \Leftrightarrow q=1+\frac{2}{\eta n}.
\end{equation}
However, only $\eta =1$ yields exact Tsallis statistics, whereas
the $p(E)$ obtained for other $\eta >0$ behave slightly different
for finite values of $E$. Asymptotically, however, only the
product of the two parameters $\eta$ and $n$ is relevant. The
same asymptotic power law can be achieved by e.g. doubling the
number of degrees of freedom $n$ if at the same time the exponent
$\eta$ is reduced by a factor $1/2$.

The case $\eta <0$ is very different. Here for small $\beta$ the
behaviour of the function $f(\beta)$ is dominated by the
exponential term
\begin{equation}
f(\beta) \sim e^{-c\beta^\eta}.
\end{equation}
According to the general results of \cite{touchette-beck}, this
implies the large energy behaviour
\begin{equation}
p(E) \sim e^{aE^{\frac{\eta}{\eta -1}}},
\end{equation}
where $a$ is a negative constant\footnote{There is a misprint in
eq.~(29) of \cite{touchette-beck}, which is hereby corrected.}
depending on $c=\frac{n}{2y_0}$ and $\eta$:
\begin{equation}
a=- \frac{1+\frac{1}{|\eta|}}{(c|\eta|)^{\frac{1}{\eta -1}}}
\end{equation}

We see that $p(E)$ now asymptotically decays as a stretched
expontial. It is interesting to see that the degrees of freedom
$n$ do not influence the exponent of the stretched exponential,
they only influence the proportionality constant $a$.

\section{Applications}

For granular gases, stretched exponential tails (and tales!) are
known to play an important role \cite{puglisi,rouyer,moon}. For
example, Rouyer and Menon \cite{rouyer} performed an experiment
where they measured the probability distribution of velocities in
a granular system confined to a vertical plane and driven by
strong vertical vibration. The system reaches a stationary state
with a probability density of horizontal velocities given by
\begin{equation}
p(v)\sim e^{-c|v|^{\alpha}},
\end{equation}
where $\alpha$ is measured to be $\alpha =1.55 \pm 0.1$. This
result is robust for all frequencies and amplitudes. The
theoretical explanation of this behaviour is still unclear.

We may assume that a superstatistics of the type described in
this paper is relevant. That means we assume that some suitable
power $y=\beta^\eta$ of the inverse granular temperature $\beta$
is $\chi^2$-distributed due to large-scale spatio-temporal
inhomogenities of granular temperature. We obtain agreement with
the experimentally observed stretched exponential tails if $\eta$
is given by
\begin{equation}
2 \frac{\eta}{\eta -1} =\alpha \Leftrightarrow \eta
=\frac{\alpha}{\alpha -2} \approx -3.4.
\end{equation}
Of course a complete theory would have to explain why in this
experiment granular temperature raised to a power of approximately
3.4 is the relevant $\chi^2$-distributed observable. Currently
this is out of reach. The value $\eta =-3.4$ seems not to be
universal. Molecular dynamics simulations \cite{moon} of dilute
granular gases driven by vertically oscillating plates yield
exponents $\alpha$ in the range $1.2 < \alpha < 1.6$, which is
equivalent to $-4.0 < \eta <-2.3$.

Stretched exponentials are sometimes also used to fit data in
hydrodynamic turbulence, despite a lack of theory for this. For
example, Bodenschatz et al. \cite{boden} present a fit of the
measured probability distributions $p(a)$ of acceleration $a$ of a
Lagrangian tracer particle which asymptotically decays as a
stretched exponential,
\begin{equation}
p(a)\sim e^{-c|a|^{0.4}}.
\end{equation}
This behaviour, interpreted in terms of a superstatistics, would
correspond to $\eta=-1/4$. This would mean that the square root of
the standard deviation of the local Gaussians would be the
relevant $\chi^2$-distributed random variable.

Overall, one has to be careful with over-ambitious fits of
experimental data using stretched exponentials. Power laws as
given by $q$-exponentials, lognormal superstatistics and stretched
exponentials can all yield very similar looking fits of
fat-tailed distributions if the largest accessible energy $E$
lies within 5-10 standard deviations. In typical experimental
situations the asymptotics of the tails is often not reached. So
there can be many competing models to explain the data.

\end{document}